\documentclass[aps,pra,superscriptaddress,twocolumn,showpacs]{revtex4}

\renewcommand*{\[}{\begin{equation}}

\renewcommand*{\]}{\end{equation}}

\def\PRA{{Phys.~Rev.~A} }

\def\JPB{{J.~Phys.~B} }
\def\PRL{{Phys.~Rev.~Lett.} }

\newcommand{\myscaleboxb}[1]{\scalebox{0.63}[0.63]{#1}}
\newcommand{\myscaleboxc}[1]{\scalebox{0.68}[0.68]{#1}}
\newcommand{\myscaleboxd}[1]{\scalebox{0.65}[0.65]{#1}}

\usepackage{epsfig}

\usepackage{hyperref}

\begin{document}

\title{Intensity dependence of multiple orbital contributions and shape resonance in high-order harmonic generation of aligned N$_{2}$ molecules}

\author{Cheng Jin}
\affiliation{J. R. Macdonald Laboratory, Physics Department, Kansas
State University, Manhattan, Kansas 66506-2604, USA}

\author{Julien B. Bertrand}
\affiliation{Joint Laboratory for Attosecond Science, National
Research Council of Canada and University of Ottawa, 100 Sussex
Drive, Ottawa, Ontario K1A 0R6, Canada}

\author{R. R. Lucchese}
\affiliation{Department of Chemistry, Texas A$\&$M University,
College Station, TX 77843-3255, USA}

\author{H. J. W\"orner}
\affiliation{Laboratorium f\"ur Physikalische Chemie, ETH Z\"urich, Wolfgang-Pauli-Strasse 10, 8093 Z\"urich, Switzerland}

\author{Paul B. Corkum}
\affiliation{Joint Laboratory for Attosecond Science, National
Research Council of Canada and University of Ottawa, 100 Sussex
Drive, Ottawa, Ontario K1A 0R6, Canada}

\author{D. M. Villeneuve}
\affiliation{Joint Laboratory for Attosecond Science, National
Research Council of Canada and University of Ottawa, 100 Sussex
Drive, Ottawa, Ontario K1A 0R6, Canada}

\author{Anh-Thu Le}
\affiliation{J. R. Macdonald Laboratory, Physics Department, Kansas
State University, Manhattan, Kansas 66506-2604, USA}

\author{C. D. Lin}
\affiliation{J. R. Macdonald Laboratory, Physics Department, Kansas State University, Manhattan, Kansas 66506-2604, USA}

\date{\today}

\begin{abstract}
We report measurements and theoretical simulations of high-order
harmonic generation (HHG) in aligned N$_2$ molecules using a 1200-nm
intense laser field when the generating pulse is perpendicular to
the aligning one. With increasing laser intensity, the minimum in
the HHG spectra first shifts its position and then disappears.
Theoretical simulations including the macroscopic propagation
effects in the medium reproduce these observations and the
disappearance of the minimum is attributed to the additional
contribution of HHG from inner orbitals. We also predict that the
well-known shape resonance in the photoionization spectra of N$_2$
should exist in the HHG spectra. It is most clearly seen when the
generating laser is parallel to the aligning one, and disappears
gradually as the angle between the two lasers increases. No clear
evidence of this shape resonance has been reported so far when using
lasers with different wavelengths. Further experimentation is needed
to draw conclusions.
\end{abstract}


\maketitle

\section{Introduction}
Recently high-order harmonic generation (HHG) has been employed to
probe the electronic structure of molecules on an ultrafast time
scale \cite{nature-imaging,
olga-nature-2009,nature-hans-2010,Haessler-jpb-review-2011,atle2009}.
After an electron is first removed from the molecule by the infrared
(IR) laser, it may be driven back by the action of an intense laser
field to recollide with the parent ion. HHG is a process where the
returning electrons recombine with the molecular ion with the
emission of high-energy photons. In fact, the last step - -
``recombination" is the inverse of photoionization. Any spectral
features in the photoionization cross section (PICS) would thus be
embodied in the HHG spectra as well. Since molecules can be
impulsively aligned by a laser field
\cite{roscapruna01a,Seideman-rmp}, the observation of HHG from
aligned molecules further offers the opportunity for probing PICSs
from aligned molecules that are not generally possible with
synchrotron radiation experiments. In addition, the first step of
HHG is a highly nonlinear tunneling ionization process which is very
selective with respect to the ionization energy of the molecular
orbital from which the electron is removed. Thus HHG in general is
dominated by the recombination to the highest occupied molecular
orbital (HOMO). In contrast, direct photoionization of a molecule is
a linear process, where PICSs from the next or next few inner
orbitals could be of comparable importance. For example, for N$_2$,
the HOMO is a $\sigma_g$ orbital. The next more tightly bound
orbital, the HOMO-1, is a $\pi_u$ orbital with an ionization energy
1.3 eV higher. The HOMO-2, is a $\sigma_u$ orbital and is 3.2 eV
more tightly bound than the HOMO. Calculations by Lucchese {\it et
al.} \cite{Lucchese-n2-82} and experimental results \cite{exp1,exp2}
show that for photon energy from the threshold at 15 eV to about 40
eV, the PICS from the HOMO-2 is negligible, but the HOMO-1 PICS is
actually comparable with that of the HOMO. On the other hand, the
HOMO has a shape resonance that peaks near the photon energy of 29
eV. These are predictions and measurements made on randomly oriented
N$_2$ molecules. How do these features depend on the alignment of
molecules? Today the rich structure in PICSs from fixed-in-space
molecules predicted by the theory remains mostly unexplored
experimentally. Can high-harmonic spectra generated by laser pulses
from aligned molecules provide new information that are not yet
directly available from photoionization measurements? The answer is
yes. Already experiments \cite{Vozzi-prl-2005,Torres-pra-2010} have
shown the existence of minimum in the HHG spectra from aligned
molecules which has not been seen in photoionization experiment.
Moreover, McFarland {\it et al.} \cite{stanford-Sci-N2} reported
that HHG from the HOMO-1 dominates over that from the HOMO when the
molecules are perpendicularly aligned with respect to the laser
polarization. Mairesse {\it et al.} \cite{Mairesse-prl-2010}
performed harmonic spectroscopy to characterize the attosecond
dynamics of multielectron rearrangement during strong-field
ionization.

Earlier attempts to extract structure information from HHG spectra
were based on the two-center interference model \cite{Lein-prl-2002}
and the strong-field approximation (SFA)
\cite{Lewen-pra-1994,Zhou-pra-2005a,Zhou-pra-2005b}. Subsequently, a
quantitative rescattering (QRS) theory
\cite{lin-jpb-10,toru-2008,at-pra-2009} was developed which
established that HHG spectra from an isolated molecule can be
expressed as the product of a returning electron wave packet with a
photo-recombination cross section (PRCS). The QRS asserts that the
PRCS is independent of the laser parameters, including the
wavelength, intensity, and pulse duration. The latter affect the
returning electron wave packet only. In the meanwhile, Smirnova {\it
et al.} \cite{olga-nature-2009} studied HHG from aligned CO$_2$
molecules and emphasized the importance of hole dynamics with
including multiple orbitals. Within the QRS theory, multiple
orbitals can be easily incorporated into the theory and it was first
used by Le {\it et al.} \cite{at-jpb-2009} to explain the HHG data
by McFarland {\it et al.} \cite{stanford-Sci-N2}. These theoretical
studies were all based on single-molecule calculations. To compare
with experimental measurements, however, the effect of the
propagation of the driving laser field and the harmonics in the
medium has to be considered. Until recently, the propagation effect
has only been considered for atomic targets, but not for molecules
\cite{jin-jpb-2011,jin-pra-ar-2011,jin-co2-2011,Zhao-pra-2011}. Even
for atoms, the propagation effect is considered mostly only within
the SFA. Since experimental HHG spectra depend critically on laser
parameters, focusing conditions, gas-jet pressure, and how the
spectra are collected, these parameters have to be available in
order to simulate the experimental spectra. When the experimental
conditions are well specified, we have shown that direct comparison
with experimental HHG spectra is now possible, for example, for
N$_2$ \cite{jin-jpb-2011,jin-pra-ar-2011} and CO$_2$
\cite{jin-co2-2011}. These comparisons are also becoming more
interesting with the emergence of the use of mid-infrared lasers to
cover a broader range of photon energies.

In this combined theoretical and experimental paper, we revisit the
HHG spectra of N$_2$ that are aligned perpendicular to the laser
polarization. This was previously studied by McFarland {\it et al.}
\cite{stanford-Sci-N2} using 800-nm lasers near the cutoff region.
Here we use 1200-nm lasers. We revisit the issue about the
importance of the HOMO-1 contribution raised in Refs.
\cite{stanford-Sci-N2,at-jpb-2009,Lee-jpb-2010}. We find that HOMO-1
does not contribute to the HHG spectra at low laser intensities, but
becomes comparable to the HOMO at higher intensities. We also
examine the shape resonance in the differential PICS, and show how
they would appear in the HHG spectra of aligned N$_2$. The paper is
organized as follows. In Sec. II, information on the experimental
setup is given. In Sec. III, we briefly summarize the theoretical
methods used in the simulation. In Sec. IV, we show the comparison
between experimental and theoretical HHG spectra, the differential
PICS of the HOMO and HOMO-1, and the shape resonance in the HHG
spectra of the HOMO for aligned N$_2$. A short summary in Sec. V
concludes this paper.

\section{Experimental setup}
To perform the experiment, we use the output of a Ti:Sapphire
multi-pass laser system (32~fs, 800~nm, 50~Hz, 12~mJ per pulse), an
optical setup to both split and recombine laser pulses, and a high
harmonic chamber, described previously \cite{Mairesse-jmo-2008},
composed of a source chamber (pulsed valve, 250~$\mu$m orifice) and
an extreme ultraviolet (XUV) spectrometer.

First, we impulsively align molecules
\cite{roscapruna01a,Seideman-rmp} with a stretched non-ionizing pump
pulse ($70\pm5$~fs, 800~nm, $I_{\text align}$=5$\times
10^{13}$~W/cm$^{2}$). Second, we probe them with a delayed intense
high harmonic generation pulse (1200 nm: $40\pm5$~fs of variable
intensity). The 1200~nm pulses ($\sim1$~mJ) come from a high-energy
optical parametric amplifier (HE-TOPAS) pumped by 800~nm light
($\sim8$~mJ). In both arms of our Mach-Zehnder interferometer, the
intensity is adjusted with neutral absorption-type density filters.
Both linearly polarized pump and probe beams are focused (f=50~cm)
$\sim$~1~mm downstream in the gas jet expansion and $\sim$~2~mm
before the orifice to select only the so-called ``short"
trajectories \cite{salieres95a}.

We record high-harmonic spectra at maximal alignment at the half
revival ($\sim$~4.12~ps). Our experimental conditions (P$_{\text
back}$=2~atm., T$_{\text rot}$=30-40K, $I_{\text align}=5\times
10^{13}$~W/cm$^{2}$, and $\tau_{\text align}$=70$\pm5$~fs) suggest
we reach a maximum degree of alignment of $\langle
cos^2\theta\rangle$=0.60$\pm0.05$ based on calculations
\cite{Mairesse2010} and supported by recent supersonic gas expansion
studies in similar conditions \cite{Yoshii2009}. We vary the angle
$\alpha$ between the alignment axis and the probing field by
rotating the alignment beam polarization, both beams are linearly
polarized. This paper concentrates on the case where
$\alpha=90^\circ$.

\section{Theoretical methods}
The details of the theoretical methods used in the simulations have
been described in Refs. \cite{jin-pra-ar-2011,jin-co2-2011}.
Briefly, both the fundamental laser field and the XUV light are
modified when they co-propagate through a macroscopic medium. For
some circumstances, such as low laser intensity, low gas pressure,
and short gas medium, the effects of dispersion, Kerr nonlinearity,
and plasmas defocusing on the fundamental laser field can be
neglected \cite{jin-pra-2009,jin-pra-ar-2011}. Under these
conditions the profile of the fundamental laser field in space (in
the vacuum) can be expressed in an analytical form. If the IR laser
is considered to be a Gaussian beam, its spatial and temporal
dependence is given approximately in Ref. \cite{jin-pra-2009}. For
the propagation of the harmonics in the medium, the dispersion and
absorption effects, which are generally anisotropic in case of
aligned molecules, are not included when the pressure is low
\cite{jin-pra-ar-2011}. We only include the induced dipoles for the
harmonics generated by the IR laser. The harmonics emitted from the
gas medium propagate in the vacuum after the gas jet, then they are
detected by the spectrometer. A Hankel transformation
\cite{far-field} is applied to obtain the far-field harmonics using
the harmonics at the exit of the gas medium.

The three-dimensional Maxwell's wave equation for the XUV light in
the gas medium is
\cite{Mette-jpb,Priori-pra-2000,tosa-pra-2005,Geissler-prl-99,jin-pra-ar-2011}
\begin{eqnarray}
\label{harm-freq}\nabla^{2}E_{\text
h}^{\parallel}(r,z,t,\alpha)-\frac{1}{c}\frac{\partial^2 E_{\text
h}^{\parallel}(r,z,t,\alpha)}{\partial^2 t} =\mu_{0}\frac{\partial^2
P_{\text{nl}}^{\parallel}(r,z,t,\alpha)}{\partial^2 t}.
\end{eqnarray}
Here $E_{\text h}^{\parallel}(r,z,t,\alpha)$ and $P_{\text
{nl}}^{\parallel}(r,z,t,\alpha)$ are the parallel components (with
respect to the polarization direction of the generating laser) of
the electric field of the XUV light and the nonlinear polarization
caused by the IR laser, respectively. $\alpha$ is the pump-probe
angle, i.e., the angle between the aligning (pump) and the
generating (probe) laser polarizations.

The nonlinear polarization term can be expressed as
\begin{eqnarray}
\label{pola}P_{\text{nl}}^{\parallel}(r,z,t,\alpha)=[n_{0}-n_{\text
e}(r,z,t,\alpha)]D^{\parallel,\text{tot}}(r,z,t,\alpha),
\nonumber\\
\end{eqnarray}
where $n_{0}-n_{\text e}(r,z,t,\alpha)$ gives the density of the
remaining neutral molecules, and
$D^{\parallel,\text{tot}}(t,\alpha)$ is the parallel component of
the induced single-molecule dipole over a number of active electrons
(including the effects from outermost and the inner orbitals). The
QRS theory \cite{lin-jpb-10,toru-2008,at-pra-2009} is applied to
calculate the induced dipole $D^{\parallel,\text{tot}}(t,\alpha)$
from each orbital. Within the QRS, the laser-induced dipole moment
$D(\omega,\theta)$ for a fixed-in-space molecule (the molecular axis
has an angle $\theta$ with respect to the laser polarization) is
given by
\begin{eqnarray}
\label{mol-qrs}D^{\parallel}(\omega,\theta)=N(\theta)^{1/2}
W(\omega)d^{\parallel}(\omega,\theta),
\end{eqnarray}
where $N(\theta)$ is the alignment-dependent ionization probability,
$W(\omega)$ is the recombining electron wave packet, and
$d^{\parallel}(\omega,\theta)$ is the parallel component of the
photorecombination (PR) transition dipole (complex in general).

For partially aligned molecules, we define an ionization weighted PR
transition dipole moment by
\begin{eqnarray}
\label{dip-avg}d^{\text
{avg}}(\omega,\alpha)=\int^{\pi}_{0}N(\theta)^{1/2}
d^{\parallel}(\omega,\theta)\rho(\theta,\alpha)\sin\theta d\theta,
\end{eqnarray}
where $\rho(\theta,\alpha)$ is the alignment distribution in the
probe-laser frame \cite{lein-jmo-2005}. Such a definition can be
made for each orbital from which the ionization occurs. The harmonic
signal from each orbital, after propagation, can be expressed as
\cite{jin-jpb-2011,jin-pra-2009,jin-pra-ar-2011}

\begin{eqnarray}
\label{mwp}S_{h}(\omega,\alpha)\propto
\omega^{4}|W(\omega)|^{2}|d^{\text {avg}}(\omega,\alpha)|^{2},
\end{eqnarray}
where $W(\omega)$ is the ``macroscopic wave packet" (MWP). If more
than one molecular orbital contributes to the HHG spectra, then the
transition dipole will be added up coherently. Note that in Eq.
(\ref{mol-qrs}) the $N(\theta)$ for each molecular orbital is
calculated using the MO-ADK theory.

Based on the previous studies of N$_2$ molecules, only the HOMO and
HOMO-1 orbitals, which are $\sigma$ and $\pi$ orbitals,
respectively, need to be included in the HHG calculation. We will
use either the terms HOMO and HOMO-1, or $\sigma$ and $\pi$
orbitals, interchangeably below.

\section{Results and discussion}
\subsection{Macroscopic HHG spectra: Experiment vs Theory}
\begin{figure}
\mbox{\rotatebox{270}{\myscaleboxd{
\includegraphics[trim=-8mm -3mm -4mm -3mm, clip]{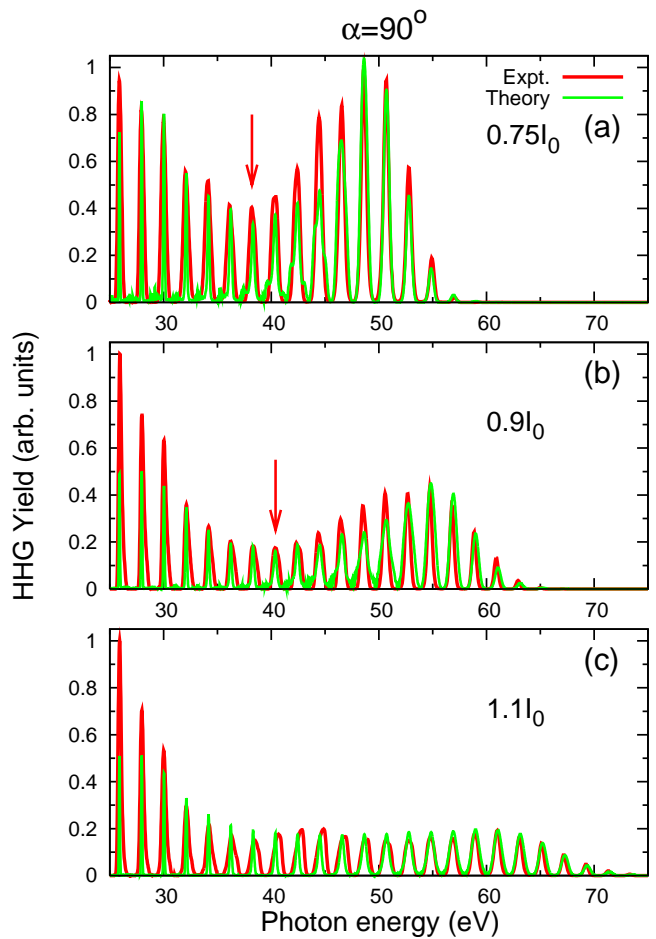}}}}
\caption{(Color online) Comparison of experimental [red (dark gray)
lines] and theoretical [green (light gray) lines] HHG spectra of
aligned N$_{2}$ in a 1200-nm generating laser where the pump-probe
angle $\alpha$=90$^{\circ}$. Laser intensities in the simulations
are indicated where I$_{0}$=10$^{14}$ W/cm$^{2}$. Degree of
alignment is $\langle\cos^{2}\theta\rangle$=0.60 by the aligning
laser. See text for additional laser parameters. Only $\sigma$
orbital is included in the simulations in (a) and (b), and both
$\sigma$ and $\pi$ orbitals are included in the simulation in (c).
Arrows indicate the positions of minima. \label{Fig1}}
\end{figure}

The HHG spectra for unaligned and aligned N$_2$ and CO$_2$ molecules
using 800-nm and 1200-nm lasers have been reported recently
\cite{hans-prl-10}. Comparison with theoretically simulated spectra
including macroscopic propagation has been carried out by Jin {\it
et al.} \cite{jin-jpb-2011,jin-pra-ar-2011,jin-co2-2011}. In this
paper we present new measurements for aligned N$_2$ molecules at a
time delay corresponding to the maximal alignment while the
HHG-generating laser is perpendicular to the aligning one. The
degree of alignment is estimated to be
$\langle\cos^{2}\theta\rangle$=0.60$\pm$0.05.

Fig.~\ref{Fig1} shows the HHG spectra of aligned N$_2$ generated by
a 1200-nm laser which is perpendicular to the aligning laser.
Experimentally, the laser duration (FWHM) is $\sim$ 44 fs, the beam
waist at the focus is $\sim$ 40 $\mu$m, and the length of gas jet is
$\sim$ 1 mm, which is located 3 mm after the laser focus. A vertical
slit with a width of 100 $\mu$m is placed 24 cm after the gas jet.
The intensities (in the center of the gas jet) used in the theory
are adjusted to coincide with experimental HHG cutoff position. In
Figs.~\ref{Fig1}(a)-~\ref{Fig1}(c), the laser intensities in theory
(experiment) are 0.75 (0.65), 0.9 (1.1), and 1.1 (1.3), in units of
10$^{14}$W/cm$^{2}$, respectively. We use the degree of alignment
$\langle\cos^{2}\theta\rangle$=0.60 and keep other parameters close
to experimental ones in the simulation. The HHG spectra from the
experiment and theory are normalized at the cutoff.

The main features in the spectra are the deep minima at 38.2 eV and
40.4 eV, at the two lower intensities in Figs.~\ref{Fig1}(a) and
\ref{Fig1}(b), respectively.  The minimum disappears at the higher
intensity in Fig.~\ref{Fig1}(c). To simulate the spectra at the two
low intensities, we include the $\sigma$ orbital only. The
simulation reproduces not only the correct shape of the spectra, but
also the precise positions of the minima in the spectra in
Figs.~\ref{Fig1}(a) and \ref{Fig1}(b). For the higher intensity in
Figs.~\ref{Fig1}(c), when we only included the $\sigma$ orbital the
theory could not reproduce the correct spectral shape. It also
predicted a minimum in the spectrum which was not seen in the
experiment [see Fig.~\ref{Fig2}(a)]. We then included both $\sigma$
and $\pi$ orbitals. A very good agreement between theory and
experiment (correct shape and no minimum in the spectrum) in
Fig.~\ref{Fig1}(c) is then achieved.

\subsection{Single HOMO orbital contribution  at low laser intensity}
We now take a careful examination of the spectral features in
Figs.~\ref{Fig1}(a) and \ref{Fig1}(b). The deep minimum is related
to the $\sigma$ orbital. This minimum has been observed in many
experiments, either in unaligned or aligned N$_2$
\cite{hans-prl-10,Torres-oe-2010,McFarland-pra-2009,Farrell-cp-2009}.
The minimum shifts only slightly when the laser intensity is
changed. This behavior is similar to the well-known Cooper minimum
in Ar \cite{jin-jpb-2011,hans-prl-2009,minemoo-pra-2008}. The same
behavior with laser intensity has been observed by W\"{o}rner {\it
et al.} \cite{hans-prl-10} (see their Fig. 1) and Farrell {\it et
al.} \cite{Farrell-cp-2009} (see their Fig. 7) when
$\alpha$=0$^{\circ}$. This can be understood as proposed in Ref.
\cite{jin-jpb-2011}. When only one molecular orbital is contributing
to the HHG spectra, the harmonic yield is given in Eq. (\ref{mwp}).
In Fig.~\ref{Fig2} we show that there is a fast drop-off in the
averaged PR transition dipole (shown is the square of the magnitude
multiplied by $\omega^2$) of the $\sigma$ orbital around 38 eV [see
Fig.~\ref{Fig2}(b)], which causes the pronounced minimum in the HHG
spectra. For different laser intensities, the slope of the averaged
PR transition dipole does not change, but the MWP
($\omega^{2}|W(\omega)|^{2}$) increases monotonically with photon
energy. Since the spatial distribution of the IR laser intensity is
different when the peak intensity at the focus is increased, the
wave packet is modified differently as it propagates through the
medium (see Fig. 3(d) in \cite{jin-jpb-2011}). This makes the
minimum in the HHG spectra shift slightly with laser intensity.
Using only one molecular orbital, our theoretical simulation can
reproduce the position of the minimum close to the experimental one.

\subsection{Multiple orbital contributions (HOMO and HOMO-1) at higher laser intensity}
\begin{figure}
\mbox{\rotatebox{270}{\myscaleboxc{
\includegraphics[trim=-10mm 3mm -3mm -1mm, clip]{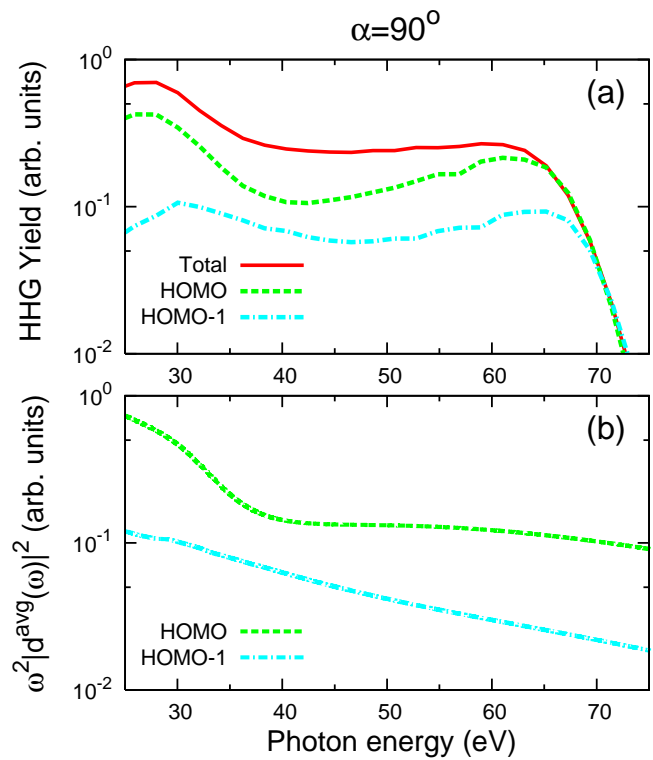}}}}
\caption{(Color online) (a) Calculated macroscopic HHG spectra
(envelope only) corresponding to Fig.~\ref{Fig1}(c). Total (HOMO and
HOMO-1 together) spectra and individual HOMO and HOMO-1 spectra. (b)
Averaged photorecombination transition dipoles (parallel component,
the square of the magnitude multiplied by $\omega^2$) of the HOMO
and HOMO-1 corresponding to (a). Laser intensity is
1.1$\times$10$^{14}$ W/cm$^{2}$. Degree of alignment is
$\langle\cos^{2}\theta\rangle$=0.60 by the aligning laser. The
pump-probe angle $\alpha$=90$^{\circ}$. \label{Fig2}}
\end{figure}

We next analyze the spectral features in Fig.~\ref{Fig1}(c). In
Fig.~\ref{Fig2}(a), the envelopes of the HHG spectra in
Fig.~\ref{Fig1}(c) from the two individual molecular orbitals, and
from the total one, are obtained after macroscopic propagation in
the gas medium, are shown. Meanwhile, the averaged PR transition
dipoles (the square of the magnitude multiplied by $\omega^2$,
degeneracy is not included) of the  two orbitals are shown in
Fig.~\ref{Fig2}(b).

In Fig.~\ref{Fig2}(a), the $\sigma$ orbital alone shows a deep
minimum, similar to Figs.~\ref{Fig1}(a) and \ref{Fig1}(b). Since the
$\pi$ orbital shows comparable contributions over the whole spectral
region, the interference between $\sigma$ and $\pi$ orbitals  washes
out the minimum in the   spectra. From Eq.~(\ref{dip-avg}), the
relative contribution between the two orbitals can be adjusted by
the alignment-dependent ionization probabilities $N(\theta)$ and the
alignment distribution $\rho(\theta,\alpha)$. For lower intensities,
the relative $N(\theta)$ of the $\pi$ orbital is small, so only the
HOMO orbital contributes to the spectra. As the intensity increases,
both orbitals contribute and the two amplitudes interfere resulting
in a drastic change of the spectra.

As shown in Eq.~(\ref{mwp}), the macroscopic HHG spectra from
individual molecular orbitals can be considered as a product of a
MWP and an averaged PR transition dipole, based on the QRS theory.
Since the ionization potential of the $\sigma$ orbital (15.6 eV)
differs from the $\pi$ orbital (16.9 eV) only by 1.3 eV, the MWPs of
the two orbitals are almost the same under the same IR laser, so the
relative contribution between $\sigma$ and $\pi$ orbitals to the
total HHG spectra is mostly determined by the averaged PR transition
dipoles.  In Figs.~\ref{Fig1}(a) and \ref{Fig1}(b), the magnitude of
$N(\theta)$ for $\sigma$ orbital is much larger than the one for the
$\pi$ orbital, thus making the corresponding averaged PR transition
dipole also larger. At the higher intensity of Fig.~\ref{Fig1}(c),
the averaged transition dipoles between the two orbitals become
comparable. Thus by increasing the laser intensity, the total HHG
can evolve from single-orbital to multiple-orbital phenomena. Note
that the MO-ADK theory \cite{tong-pra-2002,zhao-pra-10} is used to
calculate $N(\theta)$. There are also other models  in the
literature \cite{chu-pra-2009,petretti-prl-2010,at-jpb-2009} for
calculating the ionization rates. Using different ionization rates
and different alignment distributions may change the theoretical
predictions. We find that the ionization rates obtained from MO-ADK
theory is very close to the recent model calculation used by
Petretti {\it et al.} \cite{petretti-prl-2010} where they solved the
time-dependent Schr\"{o}dinger equation at laser intensity of
1.5$\times$10$^{14}$W/cm$^{2}$.

The contribution of the $\pi$ orbital to the HHG of N$_2$ when the
pump-probe angle $\alpha$=90$^\circ$ has been studied previously
\cite{stanford-Sci-N2,Lee-jpb-2010} using 800-nm pulses at the high
intensities of around 2.0$\times$10$^{14}$W/cm$^{2}$. In this case,
HOMO-1 was found to be much more pronounced in the cutoff region.
Note that the QRS theory has been applied to interpret the results
in \cite{stanford-Sci-N2} with the ionization probability obtained
from SFA \cite{at-jpb-2009} without including the propagation
effect. (Note that at the alignment angle of 90$^{\circ}$, the ratio
of ionization rate between HOMO and HOMO-1 is about 2 from the SFA
in comparison with about 5 from the MO-ADK theory
\cite{tong-pra-2002,zhao-pra-10}.) In this study, a 1200-nm laser
generates a broad photon energy range even with a low laser
intensity and the $\pi$ orbital contributes not only in the cutoff
region, but also in the plateau.

In the future, measurements similar to the present one but with a
full range of pump-probe angles may provide a way to determine the
relative ionization probabilities of the two orbitals. The
pump-probe angle $\alpha$=90$^{\circ}$ is much closer to the
alignment angle of 90$^{\circ}$ since the ``volume element" $\sin
\theta$d$\theta$ in the angular integration peaks at 90$^{\circ}$.
To probe the PR transition dipole of the HOMO orbital, HHG spectra
taken at low laser intensity with long-wavelength lasers is
preferable to avoid multiple orbital contributions. This would make
the retrieval of the structure of the target easier. If one wishes
to study the $\pi$ orbital, then a higher laser intensity and better
alignment will enhance its contribution to the HHG
\cite{stanford-Sci-N2,at-jpb-2009}.

\subsection{Photoionization amplitudes and phases of N$_2$ from HOMO and HOMO-1 orbitals}
\begin{figure*}
\mbox{\rotatebox{270}{\myscaleboxb{
\includegraphics[trim=19mm 10mm 15mm 6mm, clip]{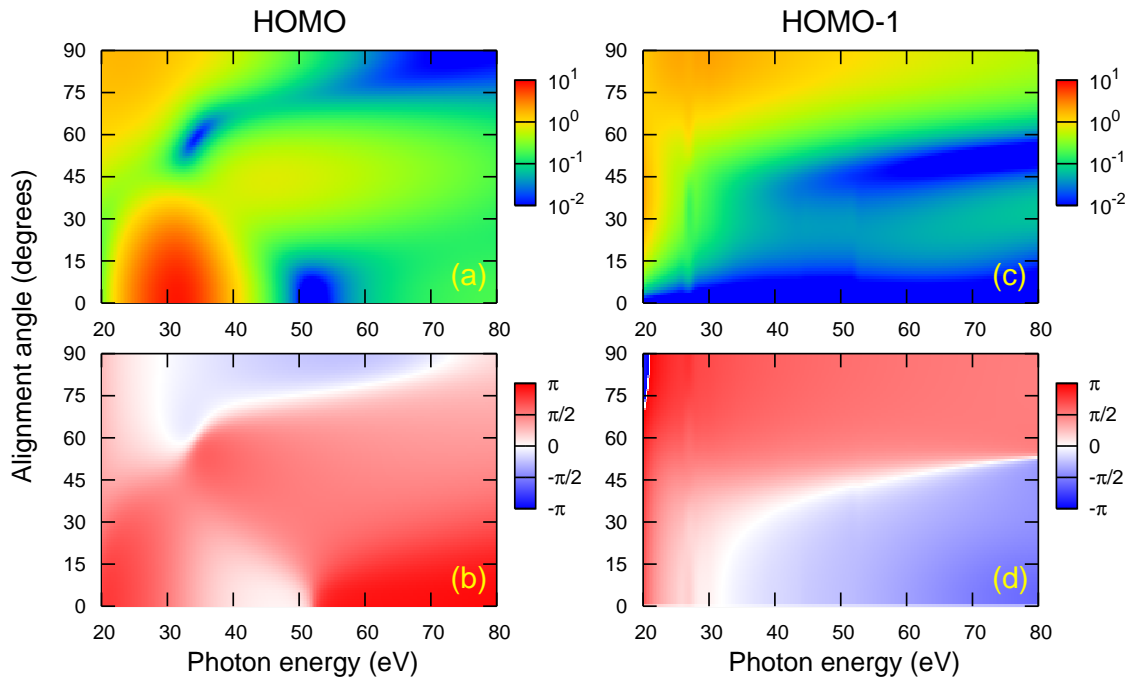}}}}
\caption{(Color online) Calculated differential photoionization cross sections [(a) and (c)] and phases [(b) and (d)] of the HOMO and HOMO-1 for N$_{2}$ in terms of the
alignment angles, respectively. The shape resonance in the HOMO shows up around 30 eV for small alignment angles only. Only the parallel component to the polarization
direction of the laser is shown. \label{Fig3}}
\end{figure*}

The most basic information on photoionization of a fixed-in-space
molecule is the parallel and perpendicular transition dipole
amplitudes in the body-frame of the molecule. They appear in the
differential PICS in the body-fixed frame as
\cite{Lucchese-n2-82,jin-pra-2010}
\begin{eqnarray}
\label{double}\frac{d^{2}\sigma}{d\Omega_{\hat{k}}d\Omega_{\hat{n}}}=\frac{4\pi^{2}\omega
k}{c} \mid \langle
\Psi_{i}|\vec{r}\cdot\hat{n}|\Psi_{f,\vec{k}}^{(-)}\rangle \mid^2,
\end{eqnarray}
where $\hat{n}$ is the polarization direction of the light,
$\vec{k}$ the momentum of the photoelectron, and $\omega$ the photon
energy. We will only focus on the case of $\hat{n}
\parallel \vec{k}$ since they are related to the parallel polarized HHG spectra measured from aligned molecules.

In Fig.~\ref{Fig3} we show the differential PICSs (in units of Mb)
and the corresponding phases of the $\sigma$ and $\pi$ orbitals of
N$_2$ using the well-established photoionization theory
\cite{Lucchese-n2-82,jin-pra-2010} for   photon energy from 20 to 80
eV. These data have been shown previously for selected photon
energies against the alignment angles for the HOMO (see Fig. 5 of
\cite{at-pra-2009}) and HOMO-1 (see Fig. 2 of \cite{at-jpb-2009}).
The observed HHG minima shown in Fig.~\ref{Fig1} above for
$\alpha$=90$^{\circ}$ and in Fig. 2 of \cite{jin-jpb-2011} for
$\alpha$=0$^{\circ}$ and for randomly distributed N$_2$ molecules
can all be understood as due to the rapid change of cross section
near 40 eV at alignment angles either close to 0$^\circ$ or
90$^\circ$. Clearly the precise position of the minimum will depend
on the degree of alignment. For the HOMO-1, Fig.~\ref{Fig3} shows
that the cross section generally peaks at large alignment angles.
Thus interference of HHG spectra from the $\sigma$ and $\pi$
orbitals are only observed close to $\alpha$=90$^{\circ}$. Note that
in Fig.~\ref{Fig3} the PICS for the two orbitals are shown on the
same scale. Except for the HOMO shape resonance near 30 eV  that
will be discussed next, the PICS from HOMO-1 is comparable with that
from HOMO over the photon energies covered. This result has been
confirmed by the measured electron spectra in photoionization
experiments. In HHG, HOMO contribution is always dominant for the
randomly distributed N$_2$. HOMO-1 can become more important only at
large alignment angles and higher laser intensities, as discussed
earlier.

\subsection{Shape resonance in HHG of aligned N$_{2}$}
\begin{figure}
\mbox{\rotatebox{270}{\myscaleboxb{
\includegraphics[trim=2mm -3mm -0.5mm -8mm, clip]{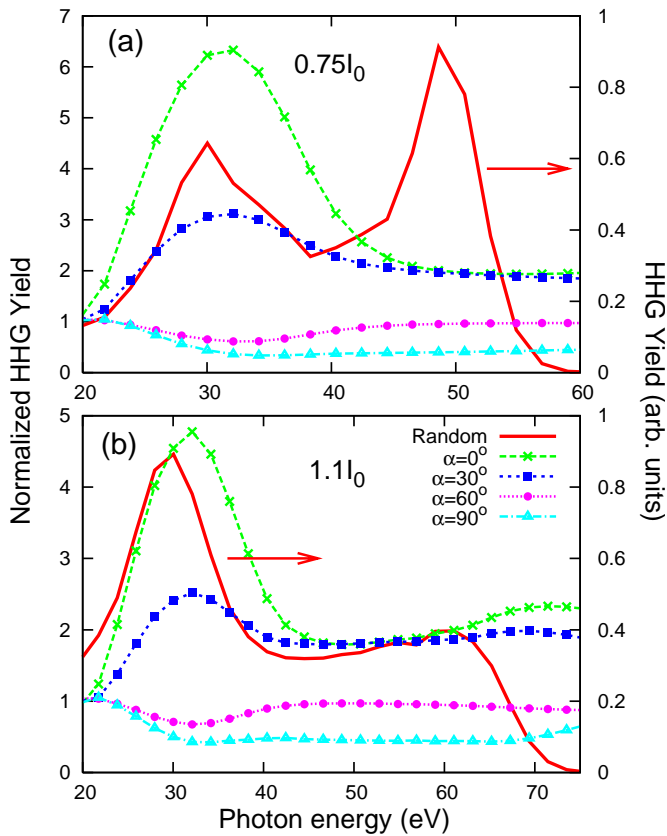}}}}
\caption{(Color online) Calculated macroscopic HHG spectra (envelope only) of unaligned N$_{2}$ and the normalized HHG yields of aligned N$_{2}$ with respect to the
unaligned ones as a function of the photon energy at the selected pump-probe angle $\alpha$. Laser intensities are indicated where I$_{0}$=10$^{14}$ W/cm$^{2}$. Degree of
alignment is $\langle\cos^{2}\theta\rangle$=0.60. \label{Fig4}}
\end{figure}

Resonances are very common in photoionization spectra. Most of them
are due to the so-called Feshbach resonances which in general are
quite narrow and thus can only be observed using high-resolution
spectroscopy. However, broad Feshbach resonances and shape
resonances can be observed since they have widths from fractions to
a few eV's. In HHG, resonances have been explored in experiment
\cite{shiner-np-2011} and theory
\cite{Strelkov-prl-2010,Lein-pra-2011,jin-Xe-2011,Frolov-pra-2011}
recently for atomic targets. But shape resonances are rare for
common atomic targets. The resonance feature observed in Xe
\cite{shiner-np-2011} is due to the intershell coupling with the
well-known shape resonance that occurs in the 4d shell. On the other
hand, shape resonances are very common in molecules. For N$_2$,
there is a pronounced shape resonance in the HOMO channel near
photon energy of 30 eV. This is due to the
3$\sigma_g$$\rightarrow$k$\sigma_u$ channel, and  for small
alignment angles only. We can see a decrease of the phase shift by
$\pi$ for this resonance from 20 to 40 eV in Fig.~\ref{Fig3}(b).
There are no other known shape resonances in the covered energy
region. Clearly this shape resonance is best observed by selecting
ionization from the $\sigma$ orbital only, by using low laser
intensity and for molecules that are aligned nearly parallel to the
polarization axis of the probing laser.

In Fig.~\ref{Fig4}, we first show the calculated HHG spectra
(envelope only, normalized) of randomly distributed N$_2$ at two
laser intensities. For randomly distributed N$_2$ molecules, it is
known that the $\sigma$ orbital is the dominant contributor to the
HHG spectra. The peak around 30 eV in the HHG spectra is due to the
shape resonance in the PICS of $\sigma$ orbital at small alignment
angles. In Fig.~\ref{Fig4}, we also show the normalized yield at
selected alignment angles with respect to the randomly distributed
one. We take the degree of alignment
$\langle\cos^{2}\theta\rangle$=0.60. The intensity of each odd
harmonic $q$ is obtained by integrating over harmonics $q-1$ to
$q+1$. For low laser intensity, the shape resonance is very
pronounced at $\alpha=0^\circ$. It decreases as the pump-probe angle
is increased, showing that the shape resonance is present only at
small aligning angles. For higher laser intensities, the same
behavior is seen even though the $\pi$ orbital is also contributing
at large pump-probe angles. We should comment that absorption is not
included in the present simulation and it will suppress the shape
resonance if the gas pressure is high.

Comparison of this prediction with existing experimental data is far
from conclusive. Torres {\it et al.} \cite{Torres-oe-2010} have
shown high harmonics data for aligned N$_2$ using a 1300-nm laser
with intensity of 1.3$\times$10$^{14}$W/cm$^{2}$ (see their Fig. 4).
The general trend of their data is very close to our prediction, but
they used a higher intensity and a higher degree of alignment
($\langle\cos^{2}\theta\rangle$=0.66), and thus the HOMO-1 channel
may contribute to the signal at larger alignment angles. Using
800-nm lasers, Lee {\it et al.} \cite{Lee-jpb-2010} reported the HHG
ratios of aligned vs unaligned N$_2$ at the selected alignment
angles. They did not present data near the resonance region and they
used a high intensity of 2.5$\times$10$^{14}$W/cm$^{2}$ [see their
Fig. 1(c)]. Our calculation does not reproduce their measured
ratios. They also reported that the normalized HHG yields for
$\alpha$=0$^\circ$ and 90$^\circ$ crosses at about harmonic order
39. This is confirmed in our calculation (not shown). The very
recent measurement by Kato {\it et al.} \cite{Kato-pra-2011} did not
extend below 30 eV either. There are other measurements
\cite{Mairesse-jmo-2008,Haessler-NatPhys-2010,Farrell-cp-2009} using
Ar or $\alpha$=0$^\circ$ as a reference. Direct comparison with
these data are difficult. Thus, it remains to be seen if the shape
resonance in N$_2$ can be seen in the HHG spectra as predicted here.
We comment that  absorption was   not included in our propagation
simulation. Absorption may modify the prediction if the gas pressure
is too high. Experiments dedicated to address this issue would be of
interest.

\section{Conclusions and outlook}
In this paper, we have reported experimental HHG spectra of aligned
N$_2$ at a time delay of the maximal alignment using a 1200-nm laser
when the pump-probe angle is set at 90$^\circ$. The minimum of the
HHG spectra appears at about 38 - 40 eV at two low laser
intensities. It disappears at a higher laser intensity. We have
carried out theoretical simulations to understand these results, and
concluded that the minima in the HHG are associated with the
properties of the transition dipole moments for photoionization from
the HOMO. At higher intensity, the contribution from HOMO-1 becomes
important and interference between the two contributions washes out
the minimum.

We have also examined the possibility of observing the well-known
shape resonance in the photoionization of N$_2$ in the HHG spectra.
While the normalized HHG yield (with respect to randomly distributed
molecules) shows clear enhancement at small alignment angles and
shape resonance may have been seen in the 1300-nm data, no evidence
of shape resonance has been observed in the HHG spectra from 800-nm
lasers. Further experiments dedicated to resolve this issue will be
of great interest.

\section{Acknowledgments}
This work was supported in part by Chemical Sciences, Geosciences
and Biosciences Division, Office of Basic Energy Sciences, Office of
Science, U.S. Department of Energy. J. B. Bertrand was also
supported by CIPI, NSERC, and AFOSR. C. Jin and C. D. Lin thank Drs.
Gae Hwang Lee and Chang Hee Nam for communicating their experimental
results.

\end{document}